\documentclass[pra,10pt,twocolumn,aps,showpacs]{revtex4-1}
\pdfoutput=1
\usepackage{amsmath}
\usepackage{amsthm}
\usepackage{amssymb}
\usepackage[colorlinks=true,linkcolor=blue,citecolor=blue,urlcolor=blue]{hyperref}
\usepackage{txfonts}
\usepackage{graphicx}
\usepackage{bbm}
\usepackage[normalem]{ulem}
\usepackage{xcolor}

\usepackage{empheq}

\newtheorem{Theorem}{Theorem}

\renewcommand\H{\mathcal{H}}
\newcommand{\1}{\mathbbm{1}}

\newcommand{\bra}[1]{\left< #1\right|}
\newcommand{\ket}[1]{\left| #1\right>}

\newcommand{\ketbra}[2]{\big| {#1}\big> \big< {#2}  \big|}

\newcommand{\tr}[1]{\textrm{tr}\! \left( #1 \right)}

\newcommand{\trz}[2]{\textrm{tr}_{#1}\! \left( #2 \right)}

\newcommand{\ncl}{\nonumber\\}
\newcommand{\proj}[1]{\ketbra{#1}{#1}}

\usepackage{array}

\begin{document}
\title{Measurement-device-independent randomness generation with arbitrary quantum states}
\author{Felix Bischof}
\email[]{felix.bischof@hhu.de}
\author{Hermann Kampermann}
\author{Dagmar Bru\ss}
\affiliation{Institut f\"ur Theoretische Physik III, Heinrich-Heine-Universit\"at D\"usseldorf, Universit\"atsstra\ss e 1, D-40225 D\"usseldorf, Germany}

\date{\today}


\begin{abstract}
Measurements of quantum systems can be used to generate classical data that is truly unpredictable for every observer. However, this true randomness needs to be discriminated from randomness due to ignorance or lack of control of the devices. 
We analyze the randomness gain of a measurement-device-independent setup, consisting of a well-characterized source of quantum states and a completely uncharacterized and untrusted detector. Our framework generalizes previous schemes as arbitrary input states and arbitrary measurements can be analyzed. Our method is used to suggest simple and realistic implementations that yield high randomness generation rates of more than one random bit per qubit for detectors of sufficient quality.
\end{abstract}

\maketitle

\section{Introduction}
\noindent
Random numbers are a fundamental resource for many information-theoretical tasks, in particular cryptography. For any task that requires secrecy, it is important that the random numbers are unpredictable for every observer, also a potential eavesdropper - a property, which is called true randomness \cite{frauch13}, or private randomness \cite{colbeck2009quantum}. This notion crucially depends on the process creating the random numbers and its underlying physics, and not just the numbers themselves.
In the deterministic classical world, randomness is the result of ignorance and hence a subjective property, which cannot be proven for a powerful adversary. In nature however, private randomness is made possible by the intrinsic unpredictability of quantum measurements: even if the whole system is known, outcomes cannot be predicted with certainty. Yet, even in quantum mechanics, true randomness cannot be certified without further assumptions. This is because realistic settings always exhibit a mixture of true quantum randomness and classical randomness. The latter may stem from uncontrolled environmental degrees of freedom, but needs to be attributed to an eavesdropper's malicious tampering with the devices. The challenge consists of separating and quantifying these types of randomness, while keeping the assumptions experimentally viable.

The amount of certifiable randomness depends on the level of control over the devices \cite{law14}. Device-independent (DI) randomness generation protocols \cite{col11,pironio2010random,arnon2016simple} view all devices as black boxes, and certify randomness via the violation of a Bell-type inequality and thus require loophole-free Bell tests. While these have recently been demonstrated experimentally \cite{giustina2015significant,shalm2015strong}, DI randomness generation setups are far from practical.

More practical schemes are obtained by introducing additional assumptions, e.g. semi-device-independent randomness generation \cite{li2011semi,brask2016high}, or the quantum steering scenario \cite{passaro2015optimal} and others. In this work, we discuss measurement-device-independent (MDI) randomness generation, of which a particular instance was introduced by \cite{ma15} and has recently been realized in experiment \cite{nie2016experimental}. The MDI setup consists of two devices: a well-characterized state source and a completely uncharacterized detector. While previous work \cite{ma15} provides the randomness generation rate of a specific two-outcome single-qubit setup, we introduce and analyze a general framework which encompasses all MDI randomness generation setups, with an arbitrary state source and detector. This allows us to devise practical setups that yield up to twice the randomness of the previous work \cite{ma15}.

This paper is structured as follows. In Sec.~\ref{sec:mdisetup} we introduce the general MDI randomness generation protocol. In Sec.~\ref{sec:rndgen} we discuss the eavesdropper's degrees of freedom and state the optimization problem in terms of a semidefinite program. Finally, examples and practical applications of our result are provided in Sec.~\ref{sec:results}.

\section{Measurement device-independent randomness generation}\label{sec:mdisetup}

\subsection{Setup and Protocol}
\begin{figure}[h]
\includegraphics[scale=0.55]{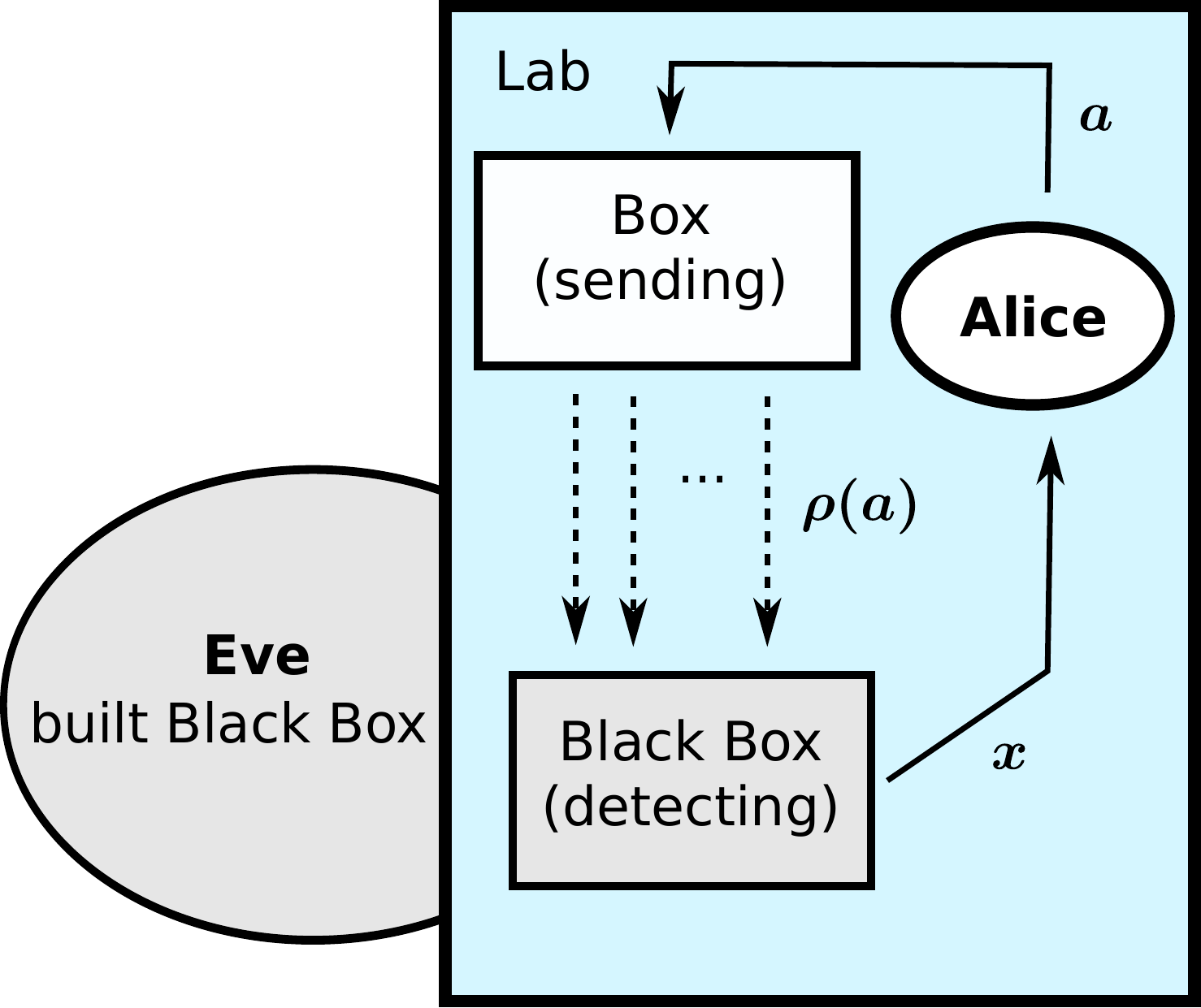} 
\caption{The measurement-device-independent setup for randomness generation (for details see text). The trusted source sends for the input $a\in\{1,\dotsc ,n_s\}$ a known state $\rho(a)$ to an untrusted measurement device (black box), which outputs $x$ with $x\in\{1,\dotsc ,n_o\}$. 
The outcome randomness $\mathcal R_X$ is characterized by the observed probability distribution $P_\textrm{obs}(x,a)$, i.e. the probability that the pair $(x,a)$ occurs.
\label{mdisetup}}
\end{figure}

The MDI randomness generation setup consists of two devices (see Fig.~\ref{mdisetup}). First,
a source, able to emit a set of well-characterized quantum states of arbitrary (finite) dimension. In particular, for the input $a$ the state $\rho(a)$ is sent. Secondly, an uncharacterized detector which announces an outcome $x$ whenever a state was sent.
The knowledge of the quantum states and the measurement results are used to characterize the detector. 

We denote the user of the protocol Alice, and the adversary Eve. Alice sets up the devices in a secure laboratory which is shielded from any kind of information transfer to the outside world.
It is verified that the sending box has no further degrees of freedom than to emit a quantum state upon receiving a specific input. The adversary Eve may have built the detector, but has no access to the laboratory afterwards.

The sent states $\rho(a)$ are selected via an initial string of random numbers.
Because of that, we describe a randomness expansion scheme \cite{colbeck2009quantum,col11}:
a user with access to an initial random string $\bold A=(a_1,a_2,\dotsc)$ interacts round-wise with a device leading to a string $\bold X=(x_1,x_2,\dotsc)$, containing the in- and output of each round, respectively. 
The randomness expansion protocol then outputs a processed string $\tilde{\bold X}(\bold X)$ which is close to uniform, conditioned on $\bold A$ as well as on any side information $E$ previously stored in the device.

The MDI randomness expansion protocol is as follows:
\begin{enumerate}
\item For every round, do Steps 2-3:
\item The sending box sends a state $\rho(a)$ of dimension $d$ with randomly chosen 
$a\in\{1,\dotsc,n_s\}$ to the measurement box. On average, this uses up $\sum_{a}p_{a}(-\log_2(p_{a}))$ bits of the initial randomness per round, where $p_{a}$ denotes the probability that $\rho(a)$ was sent.
\item After the state $\rho(a)$ has been sent, the measurement box outputs $x\in\{1,\dotsc,n_o\}$ distributed according to $p_x$. Potential losses can be announced as an extra no-detection outcome which is appended to the proper outcomes, or the device randomly attributes measurement results which contributes to the noise.
The only requirement is that the detector gives an outcome in \textit{every} round.
\item After many rounds, Alice estimates the observed measurement statistics $P_\textrm{obs}(x,a)=p_{a}P_\textrm{obs}(x|a)$, i.e. the probability that the pair $(x,a)$ occurs. From that the randomness gain per round $\mathcal R_X$ can be computed (see below).
\item Alice uses some further bits of the initial random string to post-process the raw output into a shorter string of fresh private random numbers. 
\end{enumerate}
In the last step of the protocol, the user applies a quantum-secure extraction protocol to transform the output string $\bold X$ to a string $\tilde{\bold X}$ that is close to uniform with respect to Eve and the input. This can be done via seeded extraction, e.g. two-universal hashing, for which some further random bits are needed. For details, see \cite{frauch13} and references therein.

\subsection{Randomness quantification}
\noindent
For the extraction protocol it is necessary to quantify the minimal number of bits needed for Eve to reconstruct the measurement result from her side information, i.e. the conditional min-entropy \cite{renner2008security}.
The single-round degrees of freedom in randomness expansion can be described by a tripartite state $\rho_{XAE}$ on the single-round classical out- and input registers and Eve's system, which reads
\begin{align}
\rho_{XAE}=\sum_xp_x\proj{x}_X\otimes\rho_{AE}(x) \label{cqstate},
\end{align}
where $\{\ket{x}\}$ denotes a family of orthonormal states on $X$.
The randomness contained in the random variable $X$, associated to $p_x$, is quantified by the conditional min-entropy 
\begin{align}
\mathcal R_X &= H_\textrm{min}(X|AE) 
\label{rndrate}
\end{align}
that measures the unpredictability of $X$ with respect to the classical system $A$ and the quantum system $E$. 
For $cq$-states it is known \cite{koerenn} that the min-entropy can be expressed via the optimal guessing probability 
\begin{align}
H_\textrm{min}(X|AE)= -\log_2(P^*_\textrm{guess}(X|AE)), \label{hminpguess}
\end{align}
 defined as
\begin{align}
P^*_\textrm{guess}(X|AE) = \max\limits_{\{F(x)\}}\sum_xp_x \tr{F(x)\rho_{AE}(x)}.\label{pguess}
\end{align}
Here, $\{F(x)\}$ denotes a POVM on the system $AE$, i.e. a collection of positive-semidefinite operators $F(x)\geq0$ fulfilling $\sum_{x=1}^{n_o}F(x)=\1$.

\section{Analysis of randomness generation}\label{sec:rndgen}

\subsection{Eavesdropping characterization}\label{ssec:num1}

\noindent
Before introducing the degrees of freedom in the MDI setup, we list the assumptions below.
\begin{enumerate}
\item The laboratory is shielded from any information transfer to the outside.
\item The sending device's behavior is fully characterized to emit a single specific state $\rho(a)$ upon receiving the input $a$.
\item The measurement device employs an i.i.d. strategy, i.e. it behaves independently and identically in each round.
\item We consider the asymptotic limit, i.e. the measurement statistics is precisely known.
\end{enumerate}
The first condition is necessary in any randomness expansion scheme, since otherwise the generated output could be transmitted to Eve directly. The second assumption is what differentiates MDI from fully device-independent schemes. The third condition corresponds in the language of QKD to individual attacks.
In \cite{ma15}, the authors describe how to prove security of the MDI setup against collective attacks, solely by employing the security proof against individual attacks and convexity arguments. If the arguments given there hold, this proof would also be applicable in our analysis, extending the result to collective attacks. \footnote{However, we are not sure whether the tensor product structure of the (effective) detector POVM across different rounds, employed in the proof, can be guaranteed for MDI collective attacks.}

\begin{figure}[!h]
\includegraphics[scale=0.6]{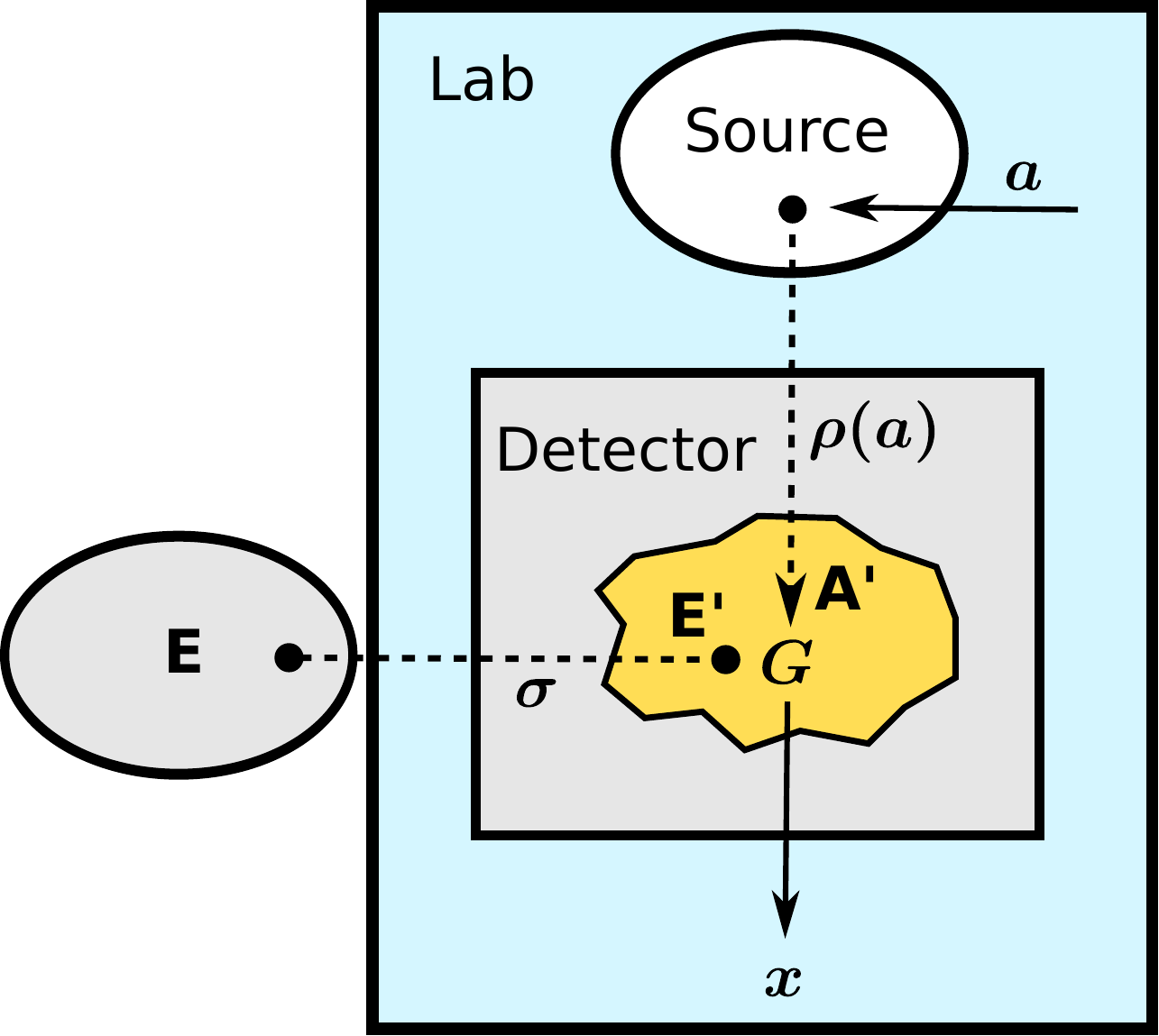}
\caption{The graphic depicts the relevant systems, states and measurement to estimate the randomness generation rate, which is conditioned on the outside systems $A$ and $E$. The primed systems are contained in the measurement box and represent internal degrees of freedom ($E'$) as well as the incoming state $\rho(a)$ ($A'$). A further internal degree of freedom is given by the unknown measurement $\{G(x)\}$, which produces the outcome $x$ in the laboratory. The state $\sigma$ provides correlations between the detector degrees of freedom and Eves site $E$.
\label{security}}
\end{figure}

Given these assumptions, the eavesdropper's most general strategy in the MDI setup is as follows. 
Eve has built the measurement apparatus that deviates from the honest device in two ways (see 
Fig.~\ref{security}). First, to obtain correlations with the measurement outcome, she has hidden a system $E'$ in the box. Her distant laboratory $E$ and the hidden system share a state $\sigma$ that may contain arbitrary amounts of entanglement.
Secondly, upon receiving the incoming states $\rho(a)$, the measurement apparatus performs an unknown measurement $\{G(x)\}$ on it and part of $\sigma$, leading to the outcome $x$ in the lab. 
Eve aims to adjust her state and the performed measurements in such a way, that she is perfectly correlated with the lab outcome, while producing the measurement statistics expected from the device.
Furthermore, the analysis includes the correlation of the output system $X$ with the input system $A$. Conditioning on the input ensures the outcome randomness to be ``fresh'', i.e. independent of the initial randomness.

\subsection{Degrees of freedom in the MDI setup}

In order to characterize the general way how the state in Eq.~(\ref{cqstate}) is obtained in the MDI setup, we introduce the relevant systems and operators below. Each system $S$ is associated to a Hilbert space $\mathcal H_S$, on which the operators act.
\begin{enumerate}
\item $A$ and $X$ denote classical registers that store the input $a$ and output $x$ of each round, respectively.
\item The incoming state $\rho(a)$ is associated to $A'$.
\item Eve has equipped the measurement apparatus with an additional system $E'$ that shares an arbitrary state $\sigma$ with her site $E$.
\item The measurement box performs an unknown POVM $\{G(x)\}$ with $n_o$ outcomes on the primed system $A'E'$ whose result is stored in $X$.
\item The optimal POVM from Eq.~(\ref{pguess}) on $AE$ is denoted by $\{F(e)\}$, with $e\in\{1,\dotsc,n_o\}$.
\end{enumerate} 
In the following, we formulate the security analysis as an optimization problem, whereby $A$ and $E$ try to guess $X$, while their operations are consistent with the classical data in the lab.
The initial average global state reads
\begin{align}
\rho_\textrm{in}=\sum_ap_a\proj{a}_A\otimes\rho_{A'}(a)\otimes\sigma_{E'E}, \label{totstate}
\end{align}
while the register $X$ is initialized with an uncorrelated state. We denote by $\rho_\textrm{in}(a)$ the initial global state if $a$ has occurred.
The box-measurement $\{G(x)\}$, which acts on $A'E'$, maps the state on $AE$ into an ensemble $\{p_x,\tau_{AE}(x)\}$ given by
\begin{align}
\tau_{AE}(x) =\frac1{p_{x}} \trz{A'E'}{G_{{A'E'}}(x)\otimes\1_{AE}\ \rho_\textrm{in}}, \label{pmstate}
\end{align}
where $\textrm{tr}_S$ denotes the partial trace over $S$.
According to Eq.~(\ref{pguess}), these states are distinguished by a measurement $\{F(e)\}$. Its outcome $e=x$ represents the system $AE$'s guess of the output $x$ of the detector.

We denote the probability of the event $(x,e)$ as $p_{x,e}$,
which is given by
\begin{align}
p_{x,e}=p_x\trz{}{F_{AE}(e)\tau_{AE}(x)}. \label{prob1}
\end{align} 
With that, the guessing probability from Eq.~(\ref{pguess}) can be formulated as $P^*_\textrm{guess}(X|AE)=\max\limits_{\{F(x)\}}\sum_x p_{x,x}$.
By combining Eq.~(\ref{pmstate}) and (\ref{prob1}), we obtain
\begin{align}
p_{x,e}
&= \trz{AE}{F_{AE}(e)\  \trz{A'E'}{G_{{A'E'}}(x)\otimes\1_{AE}\ \rho_\textrm{in}}} \ncl
&= \trz{}{G_{{A'E'}}(x)\otimes F_{AE}(e)\ \rho_\textrm{in}}, \label{prob2}
\end{align}
where in the second line we have used that 
for all linear operators $L_1$ on $\H_1$, and $\Gamma_{12}$ on $\H_1\otimes\H_2$, it holds that
$L_1\trz{2}{\Gamma_{12}}=\trz{2}{L_1\otimes\1\Gamma_{12}}$.
In Ref. \cite{tomaphd} it was proven, that when conditioning on classical information, here given by the register $A$, the optimal measurement in Eq.~(\ref{pguess}) consists of choosing an optimal POVM on $E$ for each $a$, i.e.
\begin{align}
F_{AE}(e)=\sum_{a}\proj{a}_A\otimes F_E(e|a),
\end{align}
where $\{F_E(e|a)\}$ is a family of POVMs on $E$ (with outcome $e$), indexed by $a$, fulfilling 
\begin{align}
F_E(e|a)\geq0, \quad \sum_eF_E(e|a)=\1 \qquad \forall a.\label{measnorm}
\end{align}
%
Next, we consider the action of the measurement $F_{AE}(e)$ on the initial global state $\rho_\textrm{in}(a)$ for an observer with access to $A$.
The state after measurement on $E'E$ is given by
\begin{align}
\lambda_{e|a}\sigma_{E'E}(e,a) 
= \trz{A'A}{\sqrt{F_{AE}(e)} \  \rho_\textrm{in}(a) 
	\sqrt{F_{AE}(e)}^\dagger},
\end{align}
where $\lambda_{e|a}$ denotes the probability to obtain the outcome $e$ given $a$, and $\sigma_{E'E}(e,a)$ is the corresponding conditional state. 
Note, that the unitary degree of freedom of the post-measurement state will play no role in the following, as the system $E$ will be traced out.
Since Eve's outside laboratory has no access to the input, her description of the post-measurement state is given by $\sum_ap_a\lambda_{e|a}\sigma_{E'E}(e,a)$.
Note, that because of the preparation by measurement, it holds that 
$\sigma_{E'}(e,a)=\trz{E}{\sigma_{E'E}(e,a)}$ is independent of $a$, when averaged over $e$
\begin{align}
\sum_e\lambda_{e|a}\sigma_{E'}(e,a) = \sigma_{E'} \qquad \forall a, \label{indofa}
\end{align}
i.e. the index $a$ determines a particular ensemble $\{\lambda_{e|a},\sigma_{E'}(e,a)\}$ of the state 
$\sigma_{E'}$.
This is because a local measurement, if the outcome cannot be communicated, does not influence a remote part of a state. It is known from quantum steering, that for a suitable global state, any local state can be prepared by a measurement on the other side \cite{hughston1993complete}.

Altogether, we obtain
\begin{align}\label{probb}
p_{x,e} &= \sum_{a}p_{a} p_{x,e|a} \ncl
&=\sum_{a}p_{a}\trz{}{G_{{A'E'}}(x)\rho_{A'}(a)\otimes\lambda_{e|a}\sigma_{E'}(e,a)},
\end{align}
where only the (primed) degrees of freedom in the measurement box need to be considered.

In our protocol we observe the statistics $P_\textrm{obs}(x,a)$, which constrains any valid strategy
\begin{align}
P_\textrm{obs}(x,a)&=\sum_e p_a p_{x,e|a} \ncl
&= p_{a}\trz{}{G_{{A'E'}}(x)\rho_{A'}(a)\otimes\sigma_{E'}}.
\end{align}
Here, the average over Eve's outcomes was taken, because they are unobservable for the user, and we have used Eq.~(\ref{indofa}) and (\ref{probb}) in the second line.

\subsection{The optimization problem}

We summarize the results of the previous section, by stating the optimization problem for the guessing probability. Since we are left with only two subsystems in the detector $A'E'$, we omit the system subscript
\begin{empheq}[box=\fbox]{align}
&P_\textrm{guess}^*(X|AE) = \max\limits_{\{G,\hat\sigma\}} \sum_{x,a}p_{a}\trz{}{G(x)\rho(a)\otimes\hat\sigma(x|a)} \ncl
&\textrm{such\ that}\quad G(x)\geq0, \quad \sum_x G(x)=\1, \ncl
& \hat\sigma(e|a)\geq0, \quad \sum_e\tr{\hat\sigma(e|a)}=1 \quad \forall a, \ncl
&\sum_e\hat\sigma(e|a)=\sum_e\hat\sigma(e|1) \quad \forall a, \ \textrm{and} \ncl
&P_\textrm{obs}(x,a)= \sum_e p_{a}\trz{}{G(x)\rho(a)\otimes\hat\sigma(e|a)}
 \label{bla1}
\end{empheq}
The optimization runs over ensembles $\{\hat \sigma(e|a)\}$ with $\hat \sigma(e|a)=\lambda_{e|a}\sigma(e,a)$ of arbitrary dimension and a POVM $\{G(x)\}$ acting on it and the incoming state. The fourth line represents the requirement from Eq.~($\ref{indofa}$), and the last line ensures that the detector degrees of freedom give rise to the observed probability distribution.
This optimization problem is not straightforwardly feasible, as it has a nonlinear target function with linear and semidefinite constraints.
However, we observe that the degrees of freedom relevant for the guessing probability can be combined into a single effective measurement acting only on the known state $\rho(a)$.
For that, we define an effective measurement $M_{x,e|a}$ on $\H_{A'}$ via
\begin{align}
M_{x,e|a} := \lambda_{e|a}\trz{E'}{G_{A'E'}(x)\ \1_{A'}\!\otimes\sigma_{E'}(e,a)}, \label{povm}
\end{align}
with which we can write
\begin{align}
P_\textrm{guess}^*&(X|AE) = \sum_{x,a}p_{a} \tr{M_{x,x|a}\rho_{A'}(a)}
\end{align}
by comparison with Eq.~(\ref{bla1}).
We will instead optimize over a superset of the actual degrees of freedom relevant for the guessing probability, that consists of linear operators $M_{x,e|a}$ on $\H_{A'}$ with semidefinite and linear constraints that follow from Eq.~(\ref{povm}). This, in turn, will yield an upper bound to the guessing probability and consequently a lower bound to the randomness gain.
These constraints are as follows.
The operator defined by Eq.~(\ref{povm}) is positive semidefinite, since all constituents are positive semidefinite, and furthermore fulfills for all $a$
\begin{align}
\sum_{x,e}M_{x,e|a} &= \sum_{x}\trz{E'}{G_{A'E'}(x) \1_{A'}\otimes\sigma_{E'}} = \1_{A'} ,
\end{align}
where we have used Eq.~(\ref{indofa}) in the first equality.
Thus, it has the properties of a family of POVMs on $\H_{A'}$, indexed by $a$, where the outcome $x$ goes to Alice and $e$ to Eve. Two further properties can be observed, which read
\begin{align}
&\sum_x M_{x,e|a} \propto \1, \label{povmp1}\\
&\sum_{e} M_{x,e|a}=\sum_{e} M_{x,e|a'}, \label{povmp2}
\end{align}
where Eq.~(\ref{povmp1}) follows directly from Eq.~(\ref{povm}), and Eq.~(\ref{povmp2}) follows from Eq.~(\ref{indofa}).
Thus, strategies given by a POVM family $\{M_{x,e|a}\}$ with properties (\ref{povmp1}) and (\ref{povmp2}) include the actual strategy (\ref{bla1}), but may not fully characterize it. 
The new formulation is characterized by only linear and semidefinite constraints and due to the linearity of the target function can be cast into the form of a semidefinite program (SDP).

\begin{Theorem}
The optimal guessing probability in any MDI randomness generation setup, subject to the assumptions explained in (\ref{ssec:num1}), is upper bounded by the solution of the following SDP
\end{Theorem}
\vspace*{-3ex}
\begin{empheq}[box=\fbox]{align}
&P^*_\textrm{guess}(X|AE) \leq \max\limits_{\{M_{x,e|a}\}}\sum_{x,a}p_{a} \tr{M_{x,x|a}\rho(a)}	\ncl
&\textrm{s.t.}\quad 	M_{x,e|a}\geq \,0,\quad \sum_{x,e}M_{x,e|a}= \1 \quad\forall a, \ncl
 	&\sum_x M_{x,e|a} = \left[\sum_x M_{x,e|a}\right]_{11}\cdot\1 \quad\forall e,a,\ncl
 	&\sum_{e} M_{x,e|a}=\sum_{e} M_{x,e|1}\quad\forall x,a,  \ \mbox{and}\ncl
&P_\textrm{obs}(x,a)=\sum_{e}p_{a} \tr{M_{x,e|a}\rho(a)}
\label{sdp}
\end{empheq}

The second line characterizes the operators $\{M_{x,e|a}\}$ as a POVM for each $a$. The third and fourth line ensure that the POVM family $\{M_{x,e|a}\}$ obeys the properties (\ref{povmp1}) and (\ref{povmp2}), respectively, which follow from the form of the effective measurement (\ref{povm}). The former property may be interpreted as a nonsignalling condition between the detector and Eve's site, and the latter as a nonsignalling condition between the systems $A$ and $E$.
The notation $[M]_{11}$ denotes the $(1,1)$-element of the matrix $M$. 
The last line ensures that the adversary's operations actually give rise to the observed measurement statistics $P_\textrm{obs}$ in the laboratory.
The outcome of the SDP provides via Eq.~(\ref{rndrate}) and (\ref{hminpguess}) a lower bound to the randomness generated per round $\mathcal R_X$ in the measurement-device-independent setup.

\section{Results}\label{sec:results}

For any MDI setup with arbitrary detector and state source, the observed probability vector can be read into the SDP (\ref{sdp}) to determine the randomness of the output bits.
In the following, we will discuss which sets of states $\{\rho(a)\}$ and observed distributions $p_aP_\textrm{obs}(x|a)$ are optimal in practical setups. 

Our model of the detector behavior consists of a proposed ideal quantum measurement mixed with white noise
\begin{align}
P_\textrm{obs}(x,a) &= \eta P_\textrm{id} (x,a)+\frac{1-\eta}{n_o}p_a, \label{obsdist}
\end{align}
where $P_\textrm{id}$ is the distribution of the ideal measurement, $\eta\in[0,1]$ is a quality parameter, $n_o$ is the number of outcomes, and $p_a$ is the input distribution.

In order to characterize the detector, we make use of the tomographically complete qubit state set $\{\ket{+},\ket{0},\ket{1},\ket{+i}\}$, corresponding to the $\pm1$-eigenstates of Pauli $\sigma_z$ and the $+1$-eigenstates of $\sigma_x,\sigma_y$. We employ pure states since we wish to minimize the input randomness. 

An upper bound of the MDI randomness gain is given by the classical conditional min-entropy of the in- and output distributions~\cite{tomaphd}
\begin{align}\label{clmin}
H_\textrm{min}(X|A)=-\log\sum_a\max_{x}P_\textrm{obs}(x,a).
\end{align}
In order to maximize this expression we need to have unbiased measurement outcomes $x$ for every input $a$.
Since for an ideal quantum measurement we cannot ensure unbiased outcomes with respect to each of the input states, we make use of an input distribution $p_a$ that is almost sharp, i.e. the first state $\ket{+}$ is sent with probability $q\equiv p_1\to1$, and the other states are only sent rarely to characterize the detector. We call the parameter $q$ asymmetry of the distribution $p_a$. Furthermore, an asymmetric choice of inputs is desirable for randomness expansion in the asymptotic limit, as it reduces the input randomness, see Step 2 of the protocol in Sec.~\ref{sec:mdisetup}.

To make the limit $q\to1$ feasible in the SDP, we divide all rounds into test and generation rounds: in test rounds, states are sent according to a uniform distribution, and in generation rounds, only $\rho(1)$ is sent.
The asymptotic limit is then defined as: number of rounds $N\to\infty$. Simultaneously we take the limit $q\to1$ to ensure maximal asymmetry.
Similarly to QKD Eve's optimal strategy is now as follows: 
She provides a POVM that reproduces the expected measurement statistics in the \textit{test} rounds, 
but aims at optimally predicting the outcomes of \textit{generation} rounds, since test rounds have negligible contribution to the total guessing probability in the limit $q\to1$.

In this asymmetric scenario, the optimal situation for randomness generation (expansion) corresponds to the measurement statistics of a POVM with three properties: i) the POVM is extremal \cite{d2005classical}, i.e. it 
cannot be given as a mixture of two different POVMs \cite{haapasalo2012quantum}. This ensures that its outcomes cannot be predicted by having access to a random variable (which determines the mixing) and thus maximizes randomness with respect to the measurement apparatus controlled by Eve.
ii) the POVM has unbiased outcomes for the first input state i.e. the output distribution has maximal entropy.
 iii) the POVM has $d^2$ outcomes for the state space dimension $d$. This is because $d^2$ corresponds to the highest number of independent outcomes: any further POVM element can be written as a linear combination of previous ones, which amounts to classical post-processing that cannot increase the true randomness. Therefore, the maximally achievable randomness is $2\log d$ bits \cite{acin2016optimal}.

We stress, that this POVM is realized in the optimal \textit{honest} device, i.e. a device that implements a particular pre-defined POVM.
The semidefinite program, on the other hand, finds the optimal measurement for Eve that gives rise to the measurement statistics expected from the honest device.

\subsection{Single qubit setups}

From the previous section, it follows that a qubit measurement can have up to 4 independent outcomes. In the following, we compare the performance in randomness gain of different sets of sent states and numbers of outcomes. In practice, the configuration is chosen by taking into account which states and measurements are most readily available in the laboratory.

In general, qubit POVM elements can be decomposed as
\begin{align}
M_k=\alpha_k (\1+\vec m_k \cdot \vec\sigma) \quad\textrm{with} \ncl 
\alpha_k>0,\quad \sum_k\alpha_k=1,\quad \sum_k\alpha_k\vec m_k=0. \label{extpovm}
\end{align}
where $k=1,\dotsc,n_o$.
To ensure unbiased measurement outcomes with respect to the most frequent state, we require
\begin{align}
\bra{+}M_k\ket{+}\equiv\alpha_k(1+\vec m_k\cdot \vec e_1)=\frac1{n_o} \quad
 \forall k=1,\dotsc,n_o. \label{ext}
\end{align}
Furthermore, we have the following extremality conditions \cite{d2005classical}. The POVM elements are rank-one, which is ensured by normalized measurement directions $\lvert\vec m_k \rvert=1$. Additionally, the measurement operators are linearly independent. This is fulfilled, e.g. for four outcomes, if and only if the measurement directions form a tetrahedron, i.e. they cannot lie in a common plane.

Note, that not all $\alpha_k$ can be equal, since then the property (\ref{ext}) would force all vectors to lie in the plane defined by \mbox{$\vec m_k\cdot \vec e_1=c$}, which violates the extremality condition.
A maximally symmetric 4-outcome configuration is given by 
\begin{align}\label{epovm4}
&\vec m_1=\vec e_1 \qquad \vec  m_2=-\frac17\vec e_1 + \frac{4\sqrt3}{7}\vec e_2 &\ncl
&\vec m_{3/4}=-\frac17\vec e_1 - \frac{2\sqrt3}{7}\vec e_2 \pm \frac67 \vec e_3 &\ncl
&\alpha_1=\frac18 \quad \alpha_2=\alpha_3=\alpha_4=\frac{7}{24}, &
\end{align}
which we will make use of in the following.

Also, we will later employ a 3-outcome extremal POVM given by 
\begin{align}\label{epovm3}
&\vec m_1=\vec e_2 \qquad \vec  m_{2/3}=-\frac12\vec e_2 \pm \frac{\sqrt3}{2}\vec e_3 &\ncl
&\alpha_1=\alpha_2=\alpha_3=\frac13. &
\end{align}

\begin{figure}[h!]\centering
\includegraphics[width=0.52\textwidth]{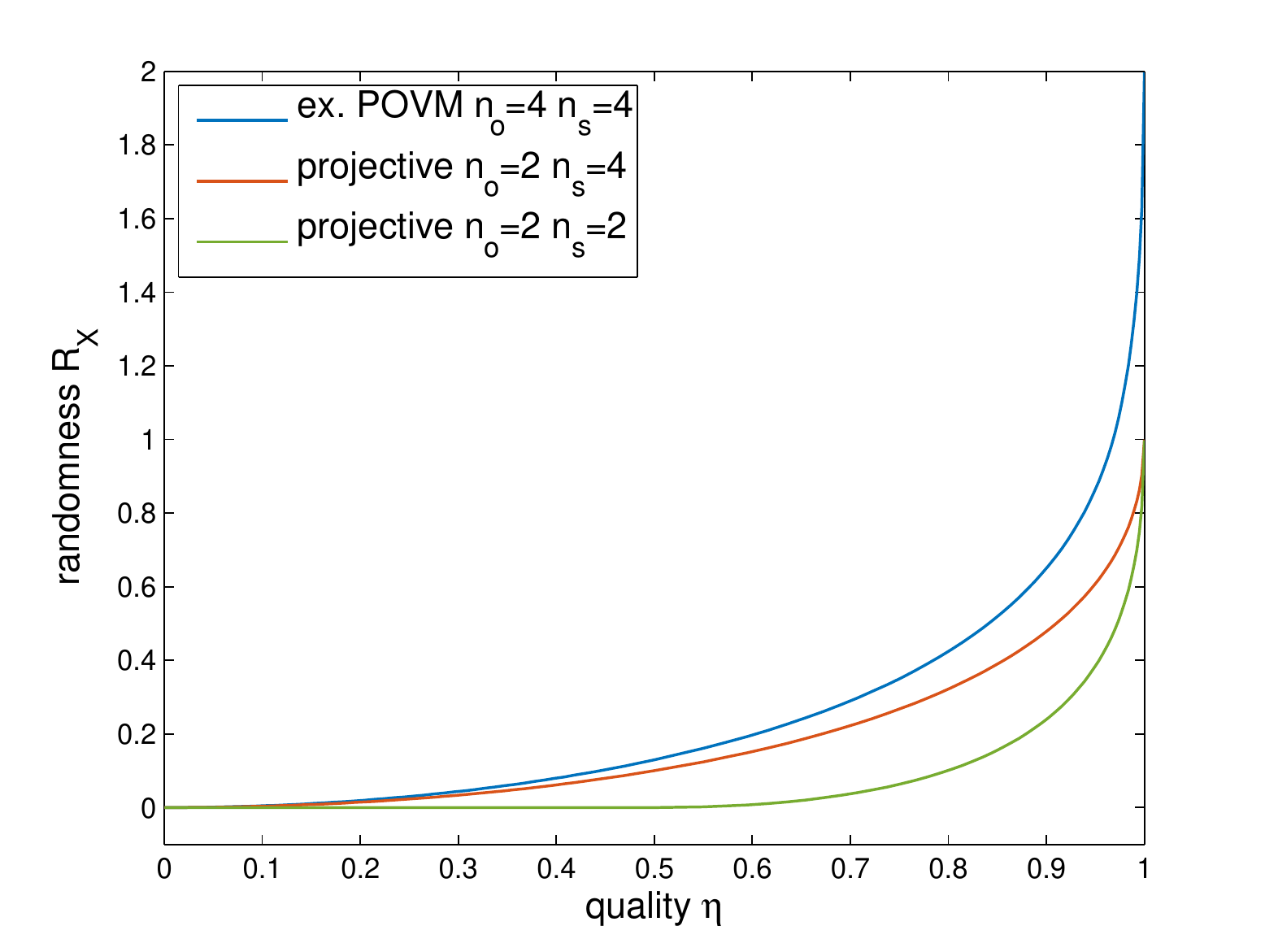}
\caption{The randomness rate versus the detector quality defined in Eq.~(\ref{obsdist}) in the asymptotic limit and for $q\equiv p_1\to1$.
The blue (red) line depicts an extremal POVM with $n_o=4\ (=2)$ outcomes for a tomographically complete set of $n_s=4$ states. The ideal measurement statistics arises from the measurement directions in Eq.~(\ref{epovm4}) ($n_o=4$) and from a $\sigma_z$-measurement ($n_o=2$). The green line corresponds to the case of two non-orthogonal sent states and two outcomes.
\label{povmvsma}}
\end{figure}

The following graphic Fig.~\ref{povmvsma} compares the performance of different numbers of outcomes and sent states in the asymptotic limit as a function of the detector quality $\eta$. For $n_s=2$ ($n_s=4$), states are drawn from the first two (all) elements of the set $\{\ket{+},\ket{0},\ket{1},\ket{+i}\}$.
The measurement statistics is described by Eq.~(\ref{obsdist}), where $P_\textrm{id}$ is the distribution, which we obtain if, for $n_o=4$, the honest device implements the 4-outcome measurement (\ref{epovm4}), and for $n_o=2$ a $\sigma_z$-measurement.
The optimization is performed with standard tools such as \texttt{YALMIP} \cite{lofberg2005yalmip}, and \texttt{SDPT3} \cite{toh1999sdpt3} as solver.
We observe that states drawn from a tomographically complete set in test rounds are clearly advantageous, since these allow for a better detector characterization.
Moreover, the figure shows that for fixed input states, the performance of an extremal 4-outcome measurement is, depending on the visibility, up to twice as good as the best projective measurement. 
In particular, the maximal local randomness of two bits is reached for a noiseless detector ($\eta=1$). For a detector quality of $\eta\geq 97\%$, the setup generates more than one random bit per qubit.
In the special case of a one-qubit sending box with tomographically complete states in the asymptotic limit, and ideal statistics given by a $\sigma_z$-measurement, our bound is equal to the exact formula from previous work \cite{ma15}. 

\subsubsection{Randomness for different relative angles}
 
\begin{figure}[h]\centering
\includegraphics[width=0.52\textwidth]{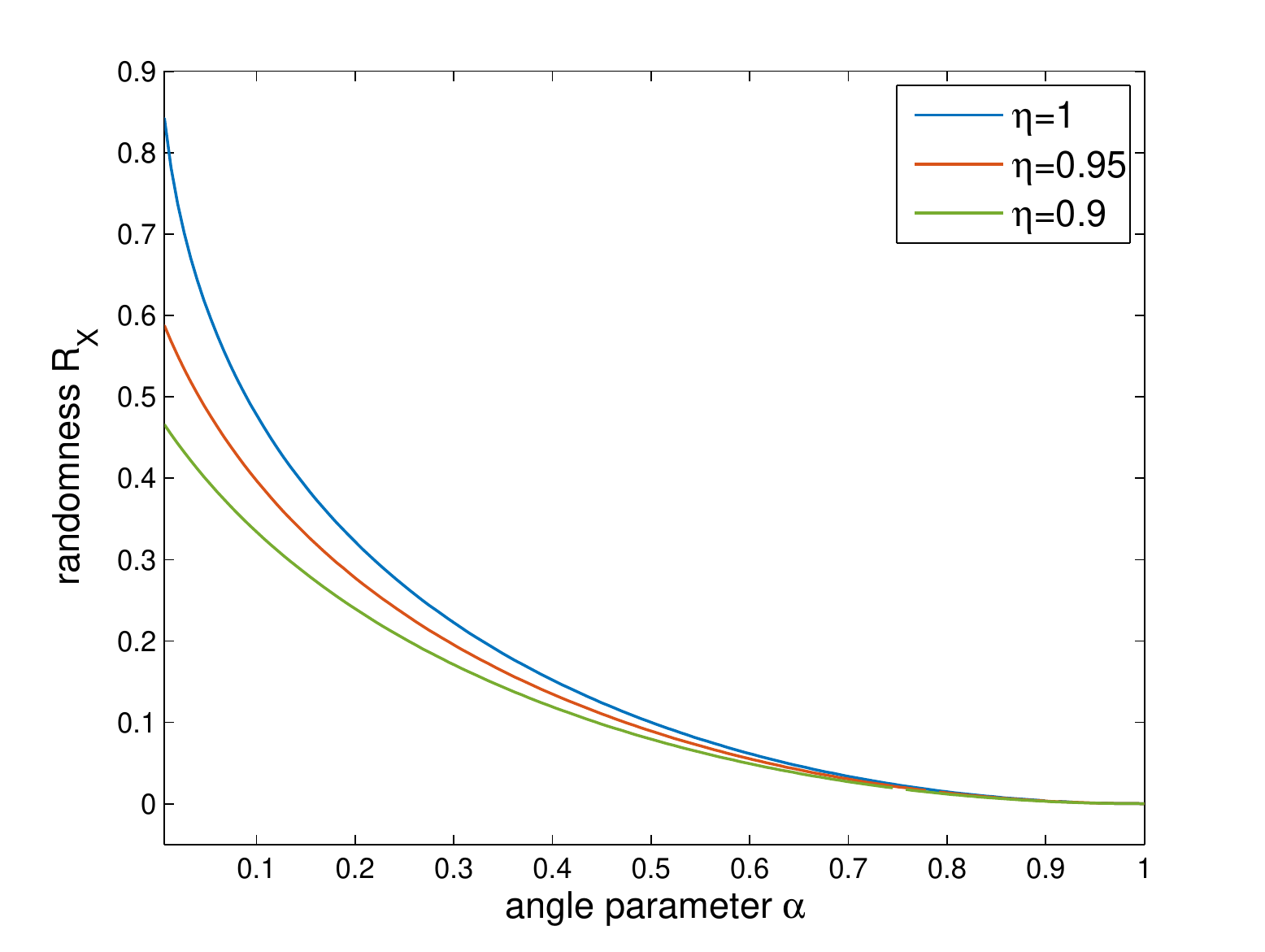}
\caption{The randomness rate as a function of the angle parameter $\alpha$ (see Eq.~(\ref{anglestates})) between two sent states for several detector qualities $\eta$ and asymmetry $q=\frac12$. The statistics of the ideal measurement corresponds to a $\sigma_x$-measurement. For $\alpha=0$, the randomness rate is equal to zero.
\label{rndangle}}
\end{figure}

We are also able to study the angle-dependency between two states of the MDI randomness rate. 
Consider the case of the observed distribution (\ref{obsdist}), where $P_\textrm{id}$ corresponds to the statistics of a $\sigma_x$-measurement. For any $\alpha \in [0,1]$, the two sent states are drawn from the set 
$\{\ket{\phi_\alpha},\ket{\psi_\alpha}\}$ with
\begin{align}\label{anglestates}
\ket{\phi_\alpha}&=\sqrt{1-\frac\alpha2}\ket{0}+\sqrt{\frac\alpha2}\ket{1}, \ncl
\ket{\psi_\alpha}&=\sqrt{1-\frac\alpha2}\ket{0}-\sqrt{\frac\alpha2}\ket{1}.
\end{align}
Fig.~\ref{rndangle} depicts the randomness generation rate for several detector qualities as a function of $\alpha \in [0,1]$. 
For $\alpha=0$ both states are identical $\ket{\phi_0}=\ket{\psi_0}=\ket{0}$, and for $\alpha=1$ they become orthogonal $\ket{\phi_1}=\ket{+},\ket{\psi_1}=\ket{-}$. In both cases the randomness generation rate vanishes, as expected. However, we observe that for an infinitesimally small but non-vanishing angle, we achieve near-maximal randomness for $\eta=1$, indicating that any amount of non-orthogonality in this scenario forces Eve to provide the honest measurement. 
More specifically, the $\eta=1$-line coincides with the classical min-entropy from Eq.~(\ref{clmin}).
However, the feature of much randomness for almost no quantumness comes at the cost of two requirements: i) precisely characterized states to ensure that they are not identical ii) precise determination of the observed measurement statistics $P_\textrm{obs}$, because the randomness rate is discontinuous at $\alpha=0$, where no randomness can be extracted.
%

\subsubsection{The role of the asymmetry}

\begin{figure}[h!]\centering
\includegraphics[width=0.52\textwidth]{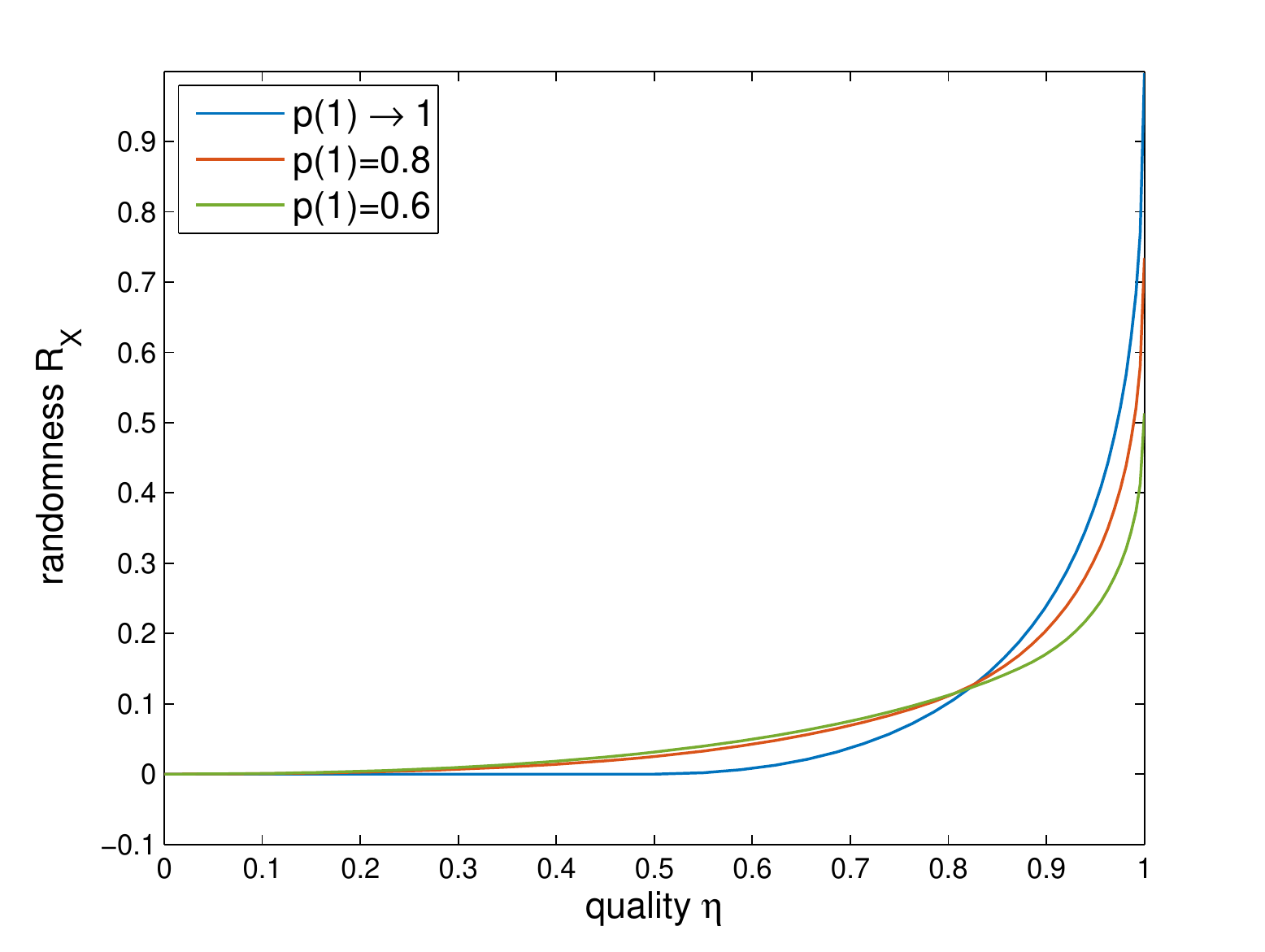}
\caption{The randomness rate of a two-state two-outcome setup with the sent states $\ket{+},\ket{0}$
 and different asymmetry parameters $q\equiv p_1$. The ideal measurement statistics corresponds to a $\sigma_z$-measurement.
\label{asymmetry}}
\end{figure}

In the asymptotic limit, and for a tomographically complete set of sent states, a higher asymmetry amounts to a higher gain in randomness for all detector qualities. However, we make the intriguing observation that for two sent qubit states $\{\ket{+},\ket{0}\}$, the optimal asymmetry depends on the detector quality. 
For that, we make use of an ideal statistics of a $\sigma_z$-measurement. Fig.~\ref{asymmetry} shows that for detector qualities $\eta\gtrsim0.8$ maximal asymmetry $q\equiv p_1\to1$ is optimal, whereas for lower qualities a more balanced input distribution performs better. 
Because $\ket{0}$ is an eigenstate of the measurement, asymmetric input distributions with higher $\ket{+}$-contribution have less classical correlation of the in- and output and thus yield higher randomness close to the ideal measurement. On the other hand, the graphic indicates that asymmetric distributions lead to higher correlations of the output and the detector degrees of freedom for an increasing noise level.

%

\subsection{Multiple qubit setups}

\subsubsection{Performance comparison}
Here, we compare setups consisting of a sending device with states of dimension $d$ and a measurement box with $n_o=d$ outcomes. In particular, the sent states are tensor products of $m$ single-qubit states which are drawn from either the first two, or all elements of the set $\{\ket{+},\ket{0},\ket{1},\ket{+i}\}$.
The measurement statistics is described by Eq.~(\ref{obsdist}), where $P_\textrm{id}$ is the distribution where the honest device implements a $\sigma_z^{\otimes m}$-measurement. 
We consider the asymmetric limit, in which the first state $\ket{+}^{\otimes m}$ is sent almost always. 
To account for experimental resources, we normalize the randomness gain to the state dimension: $\mathcal R_X/\log_2d$, which is the randomness rate per qubit.

\begin{figure}[h!]\centering
\includegraphics[width=0.52\textwidth]{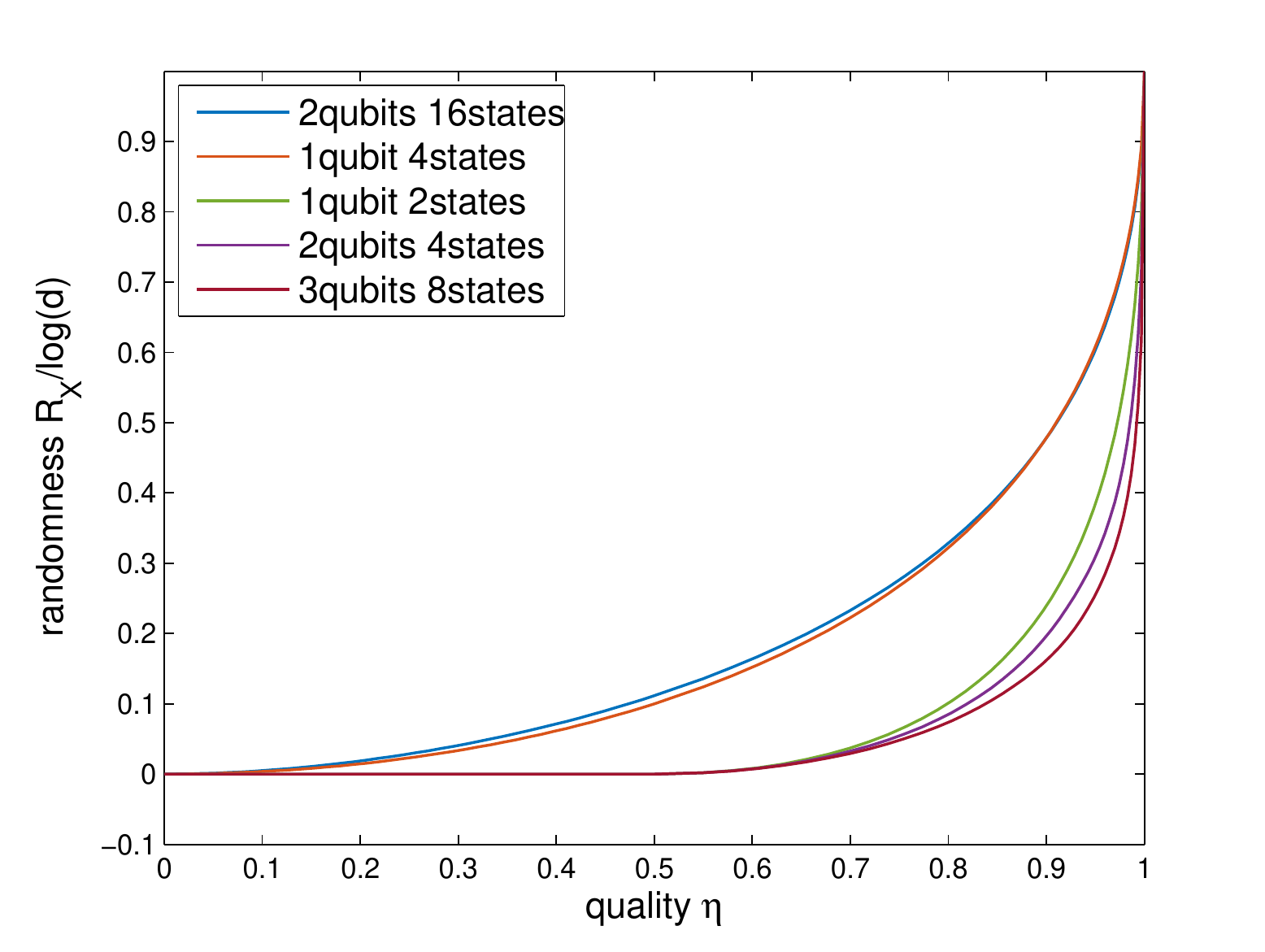}
\caption{The randomness rate per qubit versus the detector quality from Eq.~(\ref{obsdist}) for different setups in the asymptotic limit. The upper two lines correspond to setups with four different sent states per qubit, whereas the lower three lines correspond to two different sent states per qubit. The ideal measurement statistics is given by a $\sigma_z^{\otimes m}$-measurement, where $m$ is the number of sent qubits.
\label{hierarchy}}
\end{figure}

Fig.~\ref{hierarchy} depicts the randomness rate per qubit for several numbers of sent qubits $m=1,2,3$ per round. 
States drawn from a tomographically complete set in test rounds are clearly advantageous, as the upper two lines show, since these allow for a better detector characterization.
Within our noise model from Eq.~(\ref{obsdist}), the normalized randomness gain is essentially independent of the number of sent qubits. More precisely, it slightly increases with dimension for four sent states per qubit (upper two lines), and slightly decreases with dimension for two sent states per qubit (lower three lines).

%

Furthermore, we have investigated entangled measurements, such as a Bell state measurement, concluding that these do not generate more randomness and thus provide no advantage for increased experimental complexity.

\subsubsection{Individual vs. coherent attacks for two copies}
Next, we wish to compare the performance of a single qubit setup with a two-qubit setup, in which all observable quantities correspond to two independent copies of the single setup. This allows us to assess whether coherent attacks, which act simultaneously on both qubits, provide an advantage over individual attacks.
The results from Fig.~\ref{hierarchy} cannot be used for that, since there the two-qubit probability distribution is not the doubling of the one-qubit distribution.

\begin{figure}[h!]\centering
\includegraphics[width=0.52\textwidth]{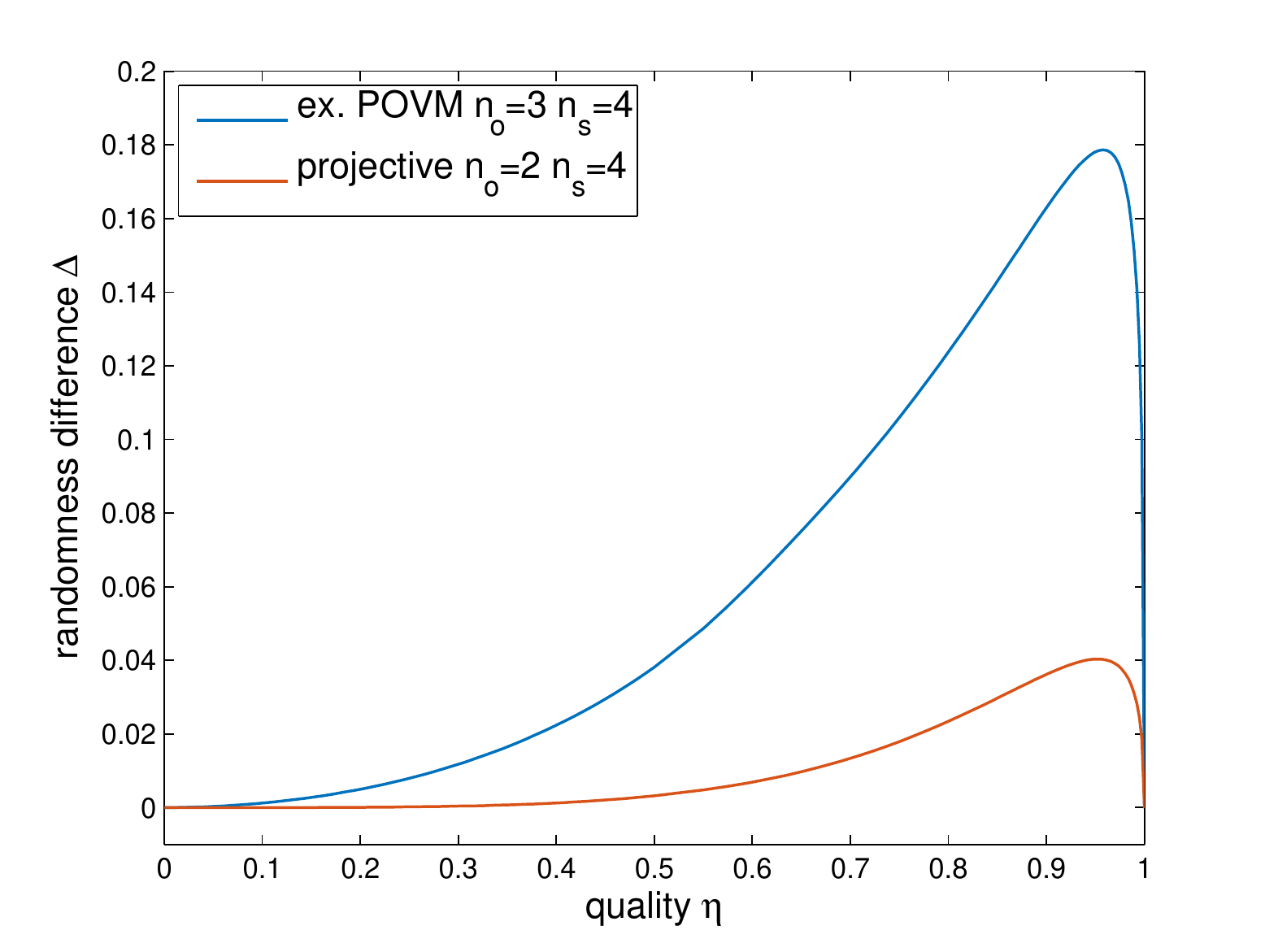}
\caption{The difference of the normalized randomness rates of a one-qubit setup with tomographically complete states, and a two-qubit doubling in the asymptotic limit. The blue line corresponds to a 3-outcome POVM, whose statistics is determined by the measurement directions in Eq.~(\ref{epovm3}), and the red line to a projective measurement.
\label{infccomp}}
\end{figure}

Fig.~\ref{infccomp} shows the difference of normalized randomness generation 
$\Delta:=\mathcal R_X(\textrm{1-qubit})-\frac12 \mathcal R_X(\textrm{2-qubit})$ of a one-qubit setup with tomographically complete states, and a two-qubit doubling. The single qubit statistics is given again by Eq.~(\ref{obsdist}).
The positive difference indicates that coherent attacks lead to more predictive power for Eve in the MDI setup. However, this assertion only holds if Eve can announce results of different round measurements simultaneously. It is an open question how Eve's predictive power behaves in a \textit{sequential} setup, where she is forced to announce an outcome in each round, but the device can have a memory. 
This means, that measurements of different rounds are in tensor product form, and act in general on the post-measurement state of all previous rounds, as well as a fresh ancilla \cite{arnon2016simple}.

\section{Conclusion and Outlook}
In this paper, we have introduced a general framework for randomness generation (i.e. expansion) with a well-characterized source of arbitrary quantum states and untrusted detector with arbitrary measurements. We presented a randomness generation protocol and analyzed its achievable gain in randomness depending on the observed measurement statistics and sent states. A lower bound on the randomness rate is calculated by a numerically feasible semidefinite program. 

As an application, we have discussed several examples of simple MDI setups and outlined optimal honest strategies. In particular, we devised a one-qubit MDI setup with four outcomes, which achieves more than one random bit per qubit for experimentally achievable detector efficiencies. 
These setups are practical compared to fully device-independent schemes, since no loophole-free Bell tests are required. Moreover, they achieve nonzero randomness generation even for low detector quality, whereas DI protocols abort in this scenario \cite{pironio2010random}.

Generalizations of our result are possible by relaxing assumptions we have made.
Of primary interest are attacks beyond the i.i.d. assumption. 
In this scenario, i.i.d. is relaxed to sequential (round-wise) interaction with the devices, including the possibility of a detector memory. For fully device-independent sequential randomness expansion, it has been shown \cite{arnon2016simple} that for more than $10^8$ rounds, the rate for general attacks is essentially the same as for i.i.d. attacks. We expect a similar behavior to hold in the case of MDI randomness expansion.
The extension to a finite number of rounds is expected to be straightforwardly implementable in the SDP in Eq.~(\ref{sdp}). In analogy to parameter estimation in QKD, one can replace equality in the last constraint by an appropriate semidefinite constraint which includes the statistical deviation.

By comparison with previous work \cite{ma15}, we noticed that for the setup treated there, our lower bound to the randomness rate coincides with their exact rate. It is an open question whether this is the case in all MDI setups, which we leave for future work.
\newline
\textit{Remark:} During completion of this work we became aware of recent related work \cite{vsupic2017measurement}, in which a comparable semidefinite program was used to calculate the MDI randomness rate in a two-qubit setup with tomographically complete states.

\section{Acknowledgments}
The authors thank Matthias Kleinmann for discussion. FB acknowledges financial support from Evangelisches Studienwerk Villigst and from Strategischer Forschungsfonds (SFF) of the University of D\"usseldorf. We acknowledge financial support from the German Federal Ministry of Education and Research (BMBF).

\bibliography{Bibliografie}{}

\begin{thebibliography}{10}

\bibitem{frauch13}
D.~Frauchiger, R.~Renner, and M.~Troyer.
\newblock True randomness from realistic quantum devices (2013).
\newblock {\em arXiv preprint arXiv:1311.4547}.

\bibitem{colbeck2009quantum}
Roger Colbeck.
\newblock Quantum and relativistic protocols for secure multi-party
  computation.
\newblock {\em arXiv preprint arXiv:0911.3814}, 2009.

\bibitem{law14}
Y.~Z. Law~et al.
\newblock Quantum randomness extraction for various levels of characterization
  of the devices.
\newblock {\em Journal of Physics A:}, 47(42):424028, 2014.

\bibitem{col11}
Roger Colbeck and Adrian Kent.
\newblock Private randomness expansion with untrusted devices.
\newblock {\em Journal of Physics A: Mathematical and Theoretical},
  44(9):095305, 2011.

\bibitem{pironio2010random}
Stefano Pironio, Antonio Ac{\'\i}n, Serge Massar, A~Boyer de~La~Giroday,
  Dzimitry~N Matsukevich, Peter Maunz, Steven Olmschenk, David Hayes, Le~Luo,
  T~Andrew Manning, et~al.
\newblock Random numbers certified by bell’s theorem.
\newblock {\em Nature}, 464(7291):1021--1024, 2010.

\bibitem{arnon2016simple}
Rotem Arnon-Friedman, Renato Renner, and Thomas Vidick.
\newblock Simple and tight device-independent security proofs.
\newblock {\em arXiv preprint arXiv:1607.01797}, 2016.

\bibitem{giustina2015significant}
Marissa Giustina, Marijn~AM Versteegh, S{\"o}ren Wengerowsky, Johannes
  Handsteiner, Armin Hochrainer, Kevin Phelan, Fabian Steinlechner, Johannes
  Kofler, Jan-{\AA}ke Larsson, Carlos Abell{\'a}n, et~al.
\newblock Significant-loophole-free test of bell’s theorem with entangled
  photons.
\newblock {\em Physical review letters}, 115(25):250401, 2015.

\bibitem{shalm2015strong}
Lynden~K Shalm, Evan Meyer-Scott, Bradley~G Christensen, Peter Bierhorst,
  Michael~A Wayne, Martin~J Stevens, Thomas Gerrits, Scott Glancy, Deny~R
  Hamel, Michael~S Allman, et~al.
\newblock Strong loophole-free test of local realism.
\newblock {\em Physical review letters}, 115(25):250402, 2015.

\bibitem{li2011semi}
Hong-Wei Li, Zhen-Qiang Yin, Yu-Chun Wu, Xu-Bo Zou, Shuang Wang, Wei Chen,
  Guang-Can Guo, and Zheng-Fu Han.
\newblock Semi-device-independent random-number expansion without entanglement.
\newblock {\em Physical Review A}, 84(3):034301, 2011.

\bibitem{brask2016high}
Jonatan~Bohr Brask, Anthony Martin, William Esposito, Raphael Houlmann, Joseph
  Bowles, Hugo Zbinden, and Nicolas Brunner.
\newblock High-rate semi-device-independent quantum random number generators
  based on unambiguous state discrimination.
\newblock {\em arXiv preprint arXiv:1612.06566}, 2016.

\bibitem{passaro2015optimal}
Elsa Passaro, Daniel Cavalcanti, Paul Skrzypczyk, and Antonio Ac{\'\i}n.
\newblock Optimal randomness certification in the quantum steering and
  prepare-and-measure scenarios.
\newblock {\em New Journal of Physics}, 17(11):113010, 2015.

\bibitem{ma15}
Z.~Cao, H.~Zhou, and X.~Ma.
\newblock Loss-tolerant measurement-device-independent quantum random number
  generation.
\newblock {\em New Journal of Physics}, 17(12):125011, 2015.

\bibitem{nie2016experimental}
You-Qi Nie, Jian-Yu Guan, Hongyi Zhou, Qiang Zhang, Xiongfeng Ma, Jun Zhang,
  and Jian-Wei Pan.
\newblock Experimental measurement-device-independent quantum random number
  generation.
\newblock {\em arXiv preprint arXiv:1612.02114}, 2016.

\bibitem{renner2008security}
Renato Renner.
\newblock Security of quantum key distribution.
\newblock {\em International Journal of Quantum Information}, 6(01):1--127,
  2008.

\bibitem{koerenn}
Robert K{\"o}nig, Renato Renner, and Christian Schaffner.
\newblock The operational meaning of min-and max-entropy.
\newblock {\em Information Theory, IEEE Transactions on}, 55(9):4337--4347,
  2009.

\bibitem{Note1}
However, we are not sure whether the tensor product structure of the
  (effective) detector POVM across different rounds, employed in the proof, can
  be guaranteed for MDI collective attacks.

\bibitem{tomaphd}
Marco Tomamichel.
\newblock A framework for non-asymptotic quantum information theory.
\newblock {\em arXiv preprint arXiv:1203.2142}, 2012.

\bibitem{hughston1993complete}
Lane~P Hughston, Richard Jozsa, and William~K Wootters.
\newblock A complete classification of quantum ensembles having a given density
  matrix.
\newblock {\em Physics Letters A}, 183(1):14--18, 1993.

\bibitem{d2005classical}
Giacomo~Mauro D'Ariano, Paoloplacido~Lo Presti, and Paolo Perinotti.
\newblock Classical randomness in quantum measurements.
\newblock {\em Journal of Physics A: Mathematical and General}, 38(26):5979,
  2005.

\bibitem{haapasalo2012quantum}
Erkka Haapasalo, Teiko Heinosaari, and Juha-Pekka Pellonp{\"a}{\"a}.
\newblock Quantum measurements on finite dimensional systems: relabeling and
  mixing.
\newblock {\em Quantum Information Processing}, 11(6):1751--1763, 2012.

\bibitem{acin2016optimal}
Antonio Ac{\'\i}n, Stefano Pironio, Tam{\'a}s V{\'e}rtesi, and Peter Wittek.
\newblock Optimal randomness certification from one entangled bit.
\newblock {\em Physical Review A}, 93(4):040102, 2016.

\bibitem{lofberg2005yalmip}
Johan Lofberg.
\newblock Yalmip: A toolbox for modeling and optimization in matlab.
\newblock In {\em Computer Aided Control Systems Design, 2004 IEEE
  International Symposium on}, pages 284--289. IEEE, 2005.

\bibitem{toh1999sdpt3}
Kim-Chuan Toh, Michael~J Todd, and Reha~H T{\"u}t{\"u}nc{\"u}.
\newblock Sdpt3—a matlab software package for semidefinite programming,
  version 1.3.
\newblock {\em Optimization methods and software}, 11(1-4):545--581, 1999.

\bibitem{vsupic2017measurement}
Ivan {\v{S}}upi{\'c}, Paul Skrzypczyk, and Daniel Cavalcanti.
\newblock Measurement-device-independent entanglement and randomness estimation
  in quantum networks.
\newblock {\em arXiv preprint arXiv:1702.04752}, 2017.

\end{thebibliography}
\bibliographystyle{unsrt}

\end{document}